\documentclass{ws-procs9x6-cpt19}
\begin{document}

\def\al{\alpha}
\def\be{\beta}
\def\ga{\gamma}
\def\de{\delta}
\def\ep{\epsilon}
\def\ve{\varepsilon}
\def\ze{\zeta}
\def\et{\eta}
\def\th{\theta}
\def\vt{\vartheta}
\def\io{\iota}
\def\vka{\varkappa}
\def\ka{\kappa}
\def\la{\lambda}
\def\vpi{\varpi}
\def\rh{\rho}
\def\vr{\varrho}
\def\si{\sigma}
\def\vs{\varsigma}
\def\ta{\tau}
\def\up{\upsilon}
\def\ph{\phi}
\def\vp{\varphi}
\def\ch{\chi}
\def\ps{\psi}
\def\om{\omega}
\def\Ga{\Gamma}
\def\De{\Delta}
\def\Th{\Theta}
\def\La{\Lambda}
\def\Si{\Sigma}
\def\Up{\Upsilon}
\def\Ph{\Phi}
\def\Ps{\Psi}
\def\Om{\Omega}
\def\cA{{\cal A}}
\def\cB{{\cal B}}
\def\cC{{\cal C}}
\def\cD{{\cal D}}
\def\cE{{\cal E}}
\def\cH{{\cal H}}
\def\cl{{\cal L}}
\def\cL{{\cal L}}
\def\cO{{\cal O}}
\def\cV{{\cal V}}
\def\cP{{\cal P}}
\def\cR{{\cal R}}
\def\cS{{\cal S}}
\def\cT{{\cal T}}

\def\fr#1#2{{{#1}\over{#2}}}
\def\frac#1#2{{\textstyle{{#1}\over{#2}}}}
\def\half{{\textstyle{1\over 2}}}
\def\ol{\overline}
\def\prt{\partial}
\def\pt{\phantom}

\def\Re{\hbox{Re}\,}
\def\Im{\hbox{Im}\,}

\def\lsim{\mathrel{\rlap{\lower4pt\hbox{\hskip1pt$\sim$}}
   \raise1pt\hbox{$<$}}}
\def\gsim{\mathrel{\rlap{\lower4pt\hbox{\hskip1pt$\sim$}}
   \raise1pt\hbox{$>$}}}

\def\etal{{\it et al.}}

\def\vev#1{\langle {#1}\rangle}
\def\expect#1{\langle{#1}\rangle}
\def\bra#1{\langle{#1}|}
\def\ket#1{|{#1}\rangle}

\def\tr{{\rm tr}}

\newcommand{\beq}{\begin{equation}}
\newcommand{\eeq}{\end{equation}}
\newcommand{\bea}{\begin{eqnarray}}
\newcommand{\eea}{\end{eqnarray}}
\newcommand{\rf}[1]{(\ref{#1})}

\def\nn{\nonumber}

\def\psb{\ol\ps{}}

\def\mbf#1{\boldsymbol #1}

\def\Q{\mathcal Q}

\def\pvec{\mbf p}
\def\gavec{\mbf\ga}

\def\Qhat{\widehat\Q}

\def\X{X}
\def\Y{Y}
\def\Z{Z}
\def\Xhat{\widehat\X}
\def\Yhat{\widehat\Y}
\def\Zhat{\widehat\Z}

\def\codt{\cos{\om_\oplus T_\oplus}}
\def\sodt{\sin{\om_\oplus T_\oplus}}
\def\ctodt{\cos{2\om_\oplus T_\oplus}}
\def\stodt{\sin{2\om_\oplus T_\oplus}}

\def\cmtemplate#1#2#3#4{{#1}^{#3}_{#4}}
\def\mfcm#1#2{\cmtemplate{m}{#1}{#2}{5}}
\def\acm#1#2{\cmtemplate{a}{#1}{#2}{}}
\def\bcm#1#2{\cmtemplate{b}{#1}{#2}{}}
\def\ccm#1#2{\cmtemplate{c}{#1}{#2}{}}
\def\dcm#1#2{\cmtemplate{d}{#1}{#2}{}}
\def\ecm#1#2{\cmtemplate{e}{#1}{#2}{}}
\def\fcm#1#2{\cmtemplate{f}{#1}{#2}{}}
\def\gcm#1#2{\cmtemplate{g}{#1}{#2}{}}
\def\Hcm#1#2{\cmtemplate{H}{#1}{#2}{}}

\def\ctemplate#1#2#3#4{{#1}^{(#2)#3}_{#4}}
\def\mc#1#2{\ctemplate{m}{#1}{#2}{}}
\def\mfc#1#2{\ctemplate{m}{#1}{#2}{5}}
\def\ac#1#2{\ctemplate{a}{#1}{#2}{}}
\def\bc#1#2{\ctemplate{b}{#1}{#2}{}}
\def\cc#1#2{\ctemplate{c}{#1}{#2}{}}
\def\dc#1#2{\ctemplate{d}{#1}{#2}{}}
\def\ec#1#2{\ctemplate{e}{#1}{#2}{}}
\def\fc#1#2{\ctemplate{f}{#1}{#2}{}}
\def\gc#1#2{\ctemplate{g}{#1}{#2}{}}
\def\Hc#1#2{\ctemplate{H}{#1}{#2}{}}

\def\mcf#1#2{\ctemplate{m}{#1}{#2}{F}}
\def\mfcf#1#2{\ctemplate{m}{#1}{#2}{5F}}
\def\acf#1#2{\ctemplate{a}{#1}{#2}{F}}
\def\bcf#1#2{\ctemplate{b}{#1}{#2}{F}}
\def\ccf#1#2{\ctemplate{c}{#1}{#2}{F}}
\def\dcf#1#2{\ctemplate{d}{#1}{#2}{F}}
\def\ecf#1#2{\ctemplate{e}{#1}{#2}{F}}
\def\fcf#1#2{\ctemplate{f}{#1}{#2}{F}}
\def\gcf#1#2{\ctemplate{g}{#1}{#2}{F}}
\def\Hcf#1#2{\ctemplate{H}{#1}{#2}{F}}

\def\mcpf#1#2{\ctemplate{m}{#1}{#2}{\prt F}}
\def\mfcpf#1#2{\ctemplate{m}{#1}{#2}{5\prt F}}
\def\acpf#1#2{\ctemplate{a}{#1}{#2}{\prt F}}
\def\bcpf#1#2{\ctemplate{b}{#1}{#2}{\prt F}}
\def\ccpf#1#2{\ctemplate{c}{#1}{#2}{\prt F}}
\def\dcpf#1#2{\ctemplate{d}{#1}{#2}{\prt F}}
\def\ecpf#1#2{\ctemplate{e}{#1}{#2}{\prt F}}
\def\fcpf#1#2{\ctemplate{f}{#1}{#2}{\prt F}}
\def\gcpf#1#2{\ctemplate{g}{#1}{#2}{\prt F}}
\def\Hcpf#1#2{\ctemplate{H}{#1}{#2}{\prt F}}

\def\cmtemplate#1#2#3#4{{#1}^{#3}_{#4}}
\def\mfcmw#1#2#3{\cmtemplate{m}{#1}{#2}{5{,#3}}}
\def\acmw#1#2#3{\cmtemplate{a}{#1}{#2}{{#3}}}
\def\bcmw#1#2#3{\cmtemplate{b}{#1}{#2}{{#3}}}
\def\ccmw#1#2#3{\cmtemplate{c}{#1}{#2}{{#3}}}
\def\dcmw#1#2#3{\cmtemplate{d}{#1}{#2}{{#3}}}
\def\ecmw#1#2#3{\cmtemplate{e}{#1}{#2}{{#3}}}
\def\fcmw#1#2#3{\cmtemplate{f}{#1}{#2}{{#3}}}
\def\gcmw#1#2#3{\cmtemplate{g}{#1}{#2}{{#3}}}
\def\Hcmw#1#2#3{\cmtemplate{H}{#1}{#2}{{#3}}}

\def\ctemplate#1#2#3#4{{#1}^{(#2)#3}_{#4}}
\def\mcw#1#2#3{\ctemplate{m}{#1}{#2}{{#3}}}
\def\mfcw#1#2#3{\ctemplate{m}{#1}{#2}{5{,#3}}}
\def\acw#1#2#3{\ctemplate{a}{#1}{#2}{{#3}}}
\def\bcw#1#2#3{\ctemplate{b}{#1}{#2}{{#3}}}
\def\ccw#1#2#3{\ctemplate{c}{#1}{#2}{{#3}}}
\def\dcw#1#2#3{\ctemplate{d}{#1}{#2}{{#3}}}
\def\ecw#1#2#3{\ctemplate{e}{#1}{#2}{{#3}}}
\def\fcw#1#2#3{\ctemplate{f}{#1}{#2}{{#3}}}
\def\gcw#1#2#3{\ctemplate{g}{#1}{#2}{{#3}}}
\def\Hcw#1#2#3{\ctemplate{H}{#1}{#2}{{#3}}}

\def\mcfw#1#2#3{\ctemplate{m}{#1}{#2}{F{,#3}}}
\def\mfcfw#1#2#3{\ctemplate{m}{#1}{#2}{5F{,#3}}}
\def\acfw#1#2#3{\ctemplate{a}{#1}{#2}{F{,#3}}}
\def\bcfw#1#2#3{\ctemplate{b}{#1}{#2}{F{,#3}}}
\def\ccfw#1#2#3{\ctemplate{c}{#1}{#2}{F{,#3}}}
\def\dcfw#1#2#3{\ctemplate{d}{#1}{#2}{F{,#3}}}
\def\ecfw#1#2#3{\ctemplate{e}{#1}{#2}{F{,#3}}}
\def\fcfw#1#2#3{\ctemplate{f}{#1}{#2}{F{,#3}}}
\def\gcfw#1#2#3{\ctemplate{g}{#1}{#2}{F{,#3}}}
\def\Hcfw#1#2#3{\ctemplate{H}{#1}{#2}{F{,#3}}}

\def\mn{{\mu\nu}}
\def\ma{{\mu\al}}
\def\mna{{\mu\nu\al}}
\def\ab{{\al\be}}
\def\bec{{\be\ga}}
\def\mab{{\mu\al\be}}
\def\mnab{{\mu\nu\al\be}}
\def\abc{{\al\be\ga}}
\def\bca{{\be\ga\al}}
\def\cab{{\ga\al\be}}
\def\mabc{{\mu\al\be\ga}}
\def\mnabc{{\mu\nu\al\be\ga}}
\def\abcd{{\al\be\ga\de}}
\def\va{{\vs\al}}
\def\vab{{\vs\al\be}}
\def\vabc{{\vs\al\be\ga}}
\def\vabcd{{\vs\al\be\ga\de}}

\def\m{m_\ps}
\def\mw{m_w}
\def\Z{\si}

\def\quar{\frac 1 4}

\def\Epn{\cE_{n,\pm1}^{e^-}}
\def\En{E_{n,\pm1}^{e^-}}
\def\app{\approx}
\def\cthodt{\cos{3\om_\oplus T_\oplus}}
\def\sthodt{\sin{3\om_\oplus T_\oplus}}
\def\note#1{{\it note \cite{#1}}}
\def\ens{E_{n,s}}
\def\enms{E_{n,-s}}
\def\at{\widetilde a}
\def\bt{\widetilde b}
\def\bft{{\widetilde b}_F}
\def\mft{{\widetilde m}_F}
\def\atw#1#2{{\widetilde a}_{#1}^{#2}}
\def\btw#1#2{{\widetilde b}_{#1}^{#2}}
\def\bftw#1#2{{\widetilde b}_{F,#1}^{#2}}
\def\mftw#1#2{{\widetilde m}_{F,#1}^{#2}}
\def\atws#1#2{{\widetilde a}_{#1}^{*#2}}
\def\btws#1#2{{\widetilde b}_{#1}^{*#2}}
\def\bftws#1#2{{\widetilde b}_{F,#1}^{*#2}}
\def\mftws#1#2{{\widetilde m}_{F,#1}^{*#2}}

\title{Comparative Penning-Trap Tests of Lorentz and CPT Symmetry}

\author{Yunhua Ding}

\address{Department of Physics, Northern Michigan University, \\
Marquette, MI 49855, USA}

\begin{abstract}
The theoretical and experimental prospects for Lorentz- and CPT-violating 
quantum electrodynamics in Penning traps are reviewed in this work.
With the recent reported results for the measurements of magnetic moments
for both protons and antiprotons,
improvements with factors of up to 3000 for the constraints of various coefficients 
for Lorentz and CPT violation are obtained. 
\end{abstract}

\bodymatter

\section{Introduction}
Among the most fundamental symmetries of relativity and particle physics
are the Lorentz and CPT invariances. 
However, in recent years it has been suggested that tiny violations of 
Lorentz and CPT symmetry are possible in a unified theory 
of gravity and quantum physics such as strings. 
\cite{string}
The comprehensive and relativistic description of such violations 
is given by the Standard-Model Extension (SME),
\cite{SME}
a general framework constructed from General Relativity and the Standard Model 
by adding to the action all possible Lorentz-violating terms.
Each of such terms is formed as the contraction of Lorentz-violating operator
with a corresponding coefficient controlling the size of Lorentz violation.
High-precision experiments across a broad range of subfields of physics,
including Penning traps, 
provide striking constraints on the coefficients for Lorentz violation. 
\cite{datatables}
In the context of minimal SME,
with operators of mass dimension restricted to $d\leq 4$,
several studies of observables for Lorentz violation in Penning traps have been conducted.
\cite{minimal}
The relevant theory of Lorentz-violating electrodynamics 
with nonminimal operators of mass dimensions up to six was also developed.
\cite{16dk}
More recently,
this treatment was generalized to include operators of arbitrary mass dimension
using gauge field theory.
\cite{19kl}
In this work, 
we further the study of experimental observables for Lorentz violation
by comparing different Penning-trap experiments
and extract new constraints on various coefficients for Lorentz violation
using the recent published results of the magnetic moments 
from the BASE collaboration.
\cite{17sc, 17sm}

\section{Theory}

For a charged Dirac fermion $\psi$ with mass $m$ confined in a Penning trap,
the magnetic moment and the related $g$-factor of the particle 
can be obtained by measuring two frequencies, 
the Larmor frequency $\nu_L=\om_L/2\pi$ and the cylotron frequency $\nu_c=\om_c/2\pi$,
and determining their ratio $g/2=\nu_L/\nu_c=\om_L/\om_c$.
In an ideal Penning-trap experiment with the magnetic field $\mbf B=B \hat x_3$
lying along the positive $x^3$ axis of the apparatus frame,
the leading-order contributions from Lorentz violation to $\om_c$ 
for both fermion and antifermions vanish,
while the corrections to $\om_L$ are given by
\bea
\de \om_L^{w} 
=
2 \btw w 3 - 2 \bftw w {33} B ,
\quad
\de \om_L^{\ol{w}}
=
- 2 \btws w 3 + 2 \bftws w {33} B ,
\label{deloma}
\eea
where $w=e^-$, $p$ for electrons and protons
and $\ol w=e^+$, $\ol p$ for positrons and antiprotons. 
The tilde quantities are defined as a combination of different coefficients for Lorentz violation.
\cite{16dk}

The expressions for the shifts in the anomaly frequencies \rf{deloma} are valid 
in the apparatus frame where the direction of the magnetic field is chosen
to be in the positive $\hat x_3$ direction.
The results in terms of the constant coefficients in the Sun-centered frame requires
a transformation matrix between the two frames,
\cite{suncenter}
which reveals the dependence of the anomaly frequencies 
on the sidereal time and the geometric conditions of the experiment.

\section{Applications}

For a confined proton or antiproton in a Penning trap,
the relevant experiment-independent observables
to the studies of the magnetic moments are the 18 coefficients for Lorentz violation
$\btw p J$, $\btws p J$, $\bftw p {(JK)}$, and $\bftws p {(JK)}$,
where $J, K=X, Y, Z$ in the Sun-centered frame.
A comparison of the magnetic moments of protons and antiprotons
in different Penning-trap experiments offers an excellent opportunity 
to constrain various of the 18 coefficients for Lorentz violation listed above.
Existing analysis involving the comparison of the magnetic moments 
of protons from the BASE collaboration at Mainz 
and of antiprotons from the ATRAP experiment at CERN
is given in our previous work.
\cite{16dk}
Recently,
the sensitivities of the magnetic moments for both protons and antiprotons
have been improved by several orders of magnitude 
by the BASE collaboration at Main and CERN.
\cite{17sc, 17sm}
Here we conduct a similar comparison analysis 
using the recent published results 
to extract improved constraints on coefficients for Lorentz violation. 

For the magnetic moment of the proton measured at Mainz,
the laboratory colatitude is $\ch \simeq 40.0^\circ$ 
and the applied magnetic field $B\simeq 1.9$ T points to local south,
which corresponds to the $\hat x$ direction in the standard laboratory frame.
For the antiproton magnetic moment measurement at CERN,
the laboratory colatitude is $\ch^* \simeq 43.8^\circ$ 
and the magnetic field $B^*\simeq 1.95$ T points $\th=60.0^\circ$ east of local north.
The experimental data for both experiments were taken over an extended time period,
so we can plausibly average the sidereal variations to be zero,
leaving only the constant parts.
Together with the numerical values of other quantities
reported by both BASE measurements, 
we obtain the bounds for $\btw p Z$, $\btws p Z$, $\bftw p {(XX)}$, $\bftws p {(XX)}$,
$\bftw p {(YY)}$, $\bftws p {(YY)}$, $\bftw p {(ZZ)}$, and $\bftws p {(ZZ)}$,
with improvement factors of up to 3000 compared to the previous results.
\cite{16dk}

Note that among the 18 independent observables 
in Penning-trap experiments
a large number of them remain unexplored to date. 
Performing a study of sidereal variations could in principle 
provide sensitivities to other components of the tilde coefficients.
Such analysis could be possible once the quantum logic readout 
currently under development at the BASE collaboration becomes feasible. 
\cite{chip}

\section*{Acknowledgments}
This work was supported in part
by the Department of Energy 
and by the Indiana University Center for Spacetime Symmetries.


\begin{thebibliography}{xx}

\bibitem{string}
V.A.~Kosteleck\'y and S.~Samuel, 
Phys.~Rev.~D  {\bf 39}, 683 (1989); 
V.A.~Kosteleck\'y and R.~Potting, 
Nucl.~Phys.~B {\bf 359}, 545 (1991); 
Phys.~Rev.~D {\bf 51}, 3923 (1995).

\bibitem{SME}
D.~Colladay and V.A.~Kosteleck\'y, 
Phys.~Rev.~D {\bf 55}, 6760 (1997); 
Phys.~Rev.~D {\bf 58}, 116002 (1998); 
V.A.~Kosteleck\'y, 
Phys.~Rev.~D {\bf 69}, 105009 (2004).

\bibitem{datatables}
{\it Data Tables for Lorentz and CPT Violation,}
V.A.~Kosteleck\'y and N.~Russell,
2019 edition,
arXiv:0801.0287v12.

\bibitem{minimal}
R.~Bluhm, V.A.~Kosteleck\'y and N.~Russell,
Phys.~Rev.~Lett. {\bf 79}, 1432 (1997); 
Phys.~Rev.~D {\bf 57}, 3923 (1998).

\bibitem{16dk}		
Y.~Ding and V.A.~Kosteleck\'y,
Phys.~Rev.~D~{\bf 94}, 056008 (2016). 

\bibitem{19kl}		
V.A.~Kosteleck\'y and Z.~Li,
Phys.~Rev.~D~{\bf 99}, 056016 (2019). 

\bibitem{17sm}
C. Smorra \etal, 
Nature~{\bf 550}, 371 (2017).
	
\bibitem{17sc}
G.~Schneider \etal,
Science~{\bf 358}, 1081 (2017).

\bibitem{suncenter}
V.A.~Kosteleck\'y and M.~Mewes, 
Phys.~Rev.~D {\bf 66}, 056005 (2002).

\bibitem{chip}
T.~Meiners \etal,
in V.A.~Kosteleck\'y, ed.,
{\it Proceedings of the Seventh Meeting on CPT and Lorentz Symmetry},
World Scientific, Singapore, 2017;
M.~Niemann \etal,
in V.A.~Kosteleck\'y, ed.,
{\it Proceedings of the Sixth Meeting on CPT and Lorentz Symmetry},
World Scientific, Singapore, 2014.

\end{thebibliography}
\end{document}